\documentclass[aps,prb,reprint,superscriptaddress,showpacs,amsmath,amssymb,twocolumn]{revtex4-2}
\usepackage{graphicx}
\usepackage{dcolumn}
\usepackage{bm}
\usepackage{physics}
\usepackage{hyperref}
\usepackage{tikz}
\usepackage{booktabs}
\usepackage{multirow}

\begin{document}

\title{Universal Transport Theory for Paired Fractional Quantum Hall States in the Quantum Point Contact Geometry}
\author{Eslam Ahmed}
\email[]{eslam.ahmed@nagoya-u.jp}
\affiliation{Department of Applied Physics, Nagoya University, Nagoya 464--8603, Japan}
\author{Ryoi Ohashi}
\email{ohashi.ryoi@mbp.phys.kyushu-u.ac.jp}
\affiliation{Department of Physics, Kyushu University, Fukuoka, 819-0395, Japan}
\author{Hiroki Isobe}
\email{isobe.hiroki@mbp.phys.kyushu-u.ac.jp}
\affiliation{Department of Physics, Kyushu University, Fukuoka, 819-0395, Japan}
\author{Kentaro Nomura}
\email{nomura.kentaro@mbp.phys.kyushu-u.ac.jp}
\affiliation{Department of Physics, Kyushu University, Fukuoka, 819-0395, Japan}
\affiliation{Quantum and Spacetime Research Institute, Kyushu University, Fukuoka 819-0395, Japan}
\author{Yukio Tanaka}
\email[]{tanaka.yukio.j2@f.mail.nagoya-u.ac.jp}
\affiliation{Department of Applied Physics, Nagoya University, Nagoya 464--8603, Japan}

\begin{abstract}
Even-denominator fractional quantum Hall (FQH) states can be viewed as topological superconductors of composite fermions, supporting a charged chiral mode and $|\mathcal{C}_{cf}|$ neutral Majorana modes set by the Chern number $\mathcal{C}_{cf}$. Despite ongoing efforts, distinguishing the many competing paired phases remains an open problem. In this work, we propose a unified theory of charge transport across a quantum point contact (QPC) for general paired FQH states described by an $so(N)_1 \times u(1)$ conformal field theory. We derive the boundary effective action for an arbitrary number of Majorana fermions $N=|\mathcal{C}_{cf}|$ and develop a non-perturbative instanton approximation to describe tunneling processes. We establish a weak-strong duality relating strong quasiparticle tunneling to weak electron tunneling. We calculate the scaling dimensions of the tunneling operators and demonstrate that while the weak-coupling fixed point is generally unstable, the strong-coupling fixed point is stable for physically relevant filling fractions and number of Majorana fermions. These transport exponents provide a distinct experimental fingerprint to identify the topological phases of even-denominator FQH states. 
\end{abstract}
\maketitle
\section{Introduction}
The nature of the half-filled fractional quantum hall (FQH) state  remains one of the most intriguing puzzles in condensed matter physics and has stayed unsolved for decades. Since the initial proposal of the Moore-Read Pfaffian state \cite{moore1991nonabelions}, which hosts non-Abelian Majorana modes obeying Ising statistics, the landscape of candidate states has expanded significantly. Competing orders include the anti-Pfaffian (aPf) state \cite{PhysRevLett.99.236806, PhysRevLett.99.236807}, the particle-hole symmetric PH-Pfaffian \cite{PhysRevX.5.031027}, the Halperin's 331 and 113 states \cite{halperin1983theory}, and the strongly paired $K=8$ state \cite{PhysRevB.61.10267}. These states are topologically distinct, characterized by different edge structures, yet they often predict similar bulk properties.

While Moore-Read's Pfaffian state was regarded as the most likely state due to early numerical studies \cite{morf1998transition,storni2010fractional,rezayi2017landau}, later studies revealed that other states such as anti-Pfaffian or PH-Pfaffian are also possible ground states \cite{PhysRevB.80.241311,PhysRevB.77.165316,PhysRevLett.117.096802}. On the other hand, when one consider multi-component 2d electron gases at half filling fractions, Halperin's 331 state and 113 state emerge as the ground state \cite{PhysRevB.82.235312,PhysRevB.81.165304}. Another possibility is the strong pairing regime of composite fermions. This regime is known as the $K=8$ state \cite{PhysRevB.61.10267}. Generally speaking, FQH states at even integer denominators are described as class D superconductors of composite fermions with Chern number $\mathcal{C}_{cf}$ \cite{PhysRevB.61.10267}. This translates into a FQH state with one charged chiral boson and $|\mathcal{C}_{cf}|$  neutral Majorana modes co-propagating (counter-propagating) on the edge when $sgn(\mathcal{C}_{cf})=1(-1)$.  Based on this theory of paired states, there is an infinite number of possible competing states at even integer denominator fillings.

Recent experiments have observed half-filled and quarter-filled fractions in 2d electron gas \cite{kumar2025quarter,chen2024tunable,singh2025fractionalquantumhallstate}. Based on the observed nearby fractions, it was argued that the observed half-filled and quarter-filled fractions alternate between Pfaffian and anti-Pfaffian states depending on the number of fully filled Landau levels \cite{singh2025fractionalquantumhallstate,kumar2025quarter,PhysRevB.110.245140,PhysRevB.110.165402,yutushui2025numericalcaseidentifyingpaired}.  The main argument is that near even denominator fractions, one should observe so-called daughter states fractions. These daughter states fractions uniquely identify the state at even denominator fractions. Note that, however, the filling fractions of daughter states can also be realized via Jain's or Halperin's sequence state which has different topological properties \cite{Singh_2024}. Thus, the nature of the state at even integer denominator is still an open problem and one needs further probes to uniquely identify it. While thermal conductance measurements can uniquely identify the topological order of FQH states \cite{Banerjee_2018}, they are prohibitively difficult to perform. Consequently, electrical transport measurements, particularly shot noise and tunneling conductance across a quantum point contact (QPC), remain indispensable tools for probing the scaling dimensions of quasiparticles directly \cite{PhysRevLett.132.136502,Hashisaka_review,10.1063/10.0034344,p4y5-trph}.

Quantum point contacts offer a means of probing edge structure. By tuning the gate voltage, one can explore a crossover between weak quasiparticle tunneling (open-QPC) and strong quasiparticle tunneling (pinched-off QPC). QPCs offer one of the earliest theoretical and experimental tools to probe FQH edges. Perhaps the earliest study involving QPCs in FQH edge dates back to the 1990s with a series of seminal works \cite{PhysRevLett.71.4381,doi:https://doi.org/10.1002/9783527617258.ch4,PhysRevB.46.7268,PhysRevB.46.15233,PhysRevB.51.2363} for the Laughlin $\nu=1/3$ state and with the first exact solution given by Ref. \cite{PhysRevLett.74.3005,PhysRevB.52.8934}. The main insight from these earlier studies is the establishment of a duality relating strong quasiparticle tunneling to weak electron tunneling using the dilute instanton approximation, allowing for perturbative treatment of strong quasiparticle tunneling.  For paired states, the interplay between charge and neutral sectors can lead to highly nontrivial scaling behavior. To our knowledge, the earliest work studying QPCs in paired states dates back to late 1990s in Ref. \cite{imura1998tunneling}, where several competing states known at that time were extensively analyzed. Here, the strong-weak duality was assumed to hold based on physical arguments. The first proof of the strong-weak duality was done by one of the authors (K. Nomura) in the Pfaffian state and Halperin's 331 state \cite{Nomura_2001} using instanton approximation. A subsequent work re-derived the duality using alternative approach \cite{PhysRevLett.97.036801}. The results by Nomura et al was generalized to the Anti-Pfaffian state shortly after its discovery \cite{Ito_2012}. The first systematic and exhaustive study of the tunneling conductance for all competing paired states was done in 2019 in the weak quasiparticle tunneling limit where perturbation theory is possible \cite{PhysRevB.100.035302}. Another related QPC problem is Andreev-like reflection between two different FQH states separated by a QPC. The earliest such proposal was studied in a QPC separating $\nu=1$ and $\nu=1/3$ FQH states where a similar strong-weak duality was derived using dilute instanton method \cite{sandlerAndreevReflectionFractional1998}. This result was generalized to a QPC separating $\nu=1$ and $\nu=1/2$ Pfaffian states by three of the authors (R. Ohashi, K. Nomura, Y. Tanaka) \cite{ohashi2022andreev}. 
While previous studies analyzed specific states (Pfaffian,\cite{PhysRevLett.97.036801,Nomura_2001} anti-Pfaffian,\cite{Ito_2012} 331,\cite{Nomura_2001} etc.), a general systematic framework for arbitrary $N$ Majorana modes has been lacking especially in both the weak and strong quasiparticle tunneling regimes. More importantly, systematic proof of the strong-weak duality is largely left unexplored.

In this paper, we develop a unified transport theory for a general class of paired FQH states described by $so(N)_1 \times u(1)$ conformal field theory (CFT) \cite{yellow_book_cft}. Here, the integer $N=|\mathcal{C}_{cf}|$ corresponds to the number of neutral Majorana fermion modes at the edge. This framework captures the Pfaffian ($N=1$), 331 ($N=2$),  anti-Pfaffian ($N=3$),  and $K=8$ ($N=0$) states among others within a single formalism. We note that this formalism matches with the sixteenfold way formalism of Ref. \cite{PhysRevB.100.035302} and are related to each other via bosonization of pairs of Majoranas in the neutral sector.

We consider a QPC geometry where the opposite edges of the FQH bar are brought into proximity. We analyze quasiparticle tunneling in the weak and strong tunneling regimes. Using an instanton expansion of the boundary partition function, we derive a duality transformation that maps the strong quasiparticle tunneling regime to the weak electron tunneling regime. We calculate the scaling dimensions of the tunneling operators for arbitrary number of Majorana modes $N$ and filling fraction $\nu$. Our results show that the scaling dimension of the quasiparticle tunneling operator is $\Delta_{qp} = \frac{2\nu+N}{8}$, rendering it relevant for typical FQH fractions ($\nu < 1$) and realistic number of neutral Majorana modes ($N<8$). This implies that the insulating state is a stable fixed point, contrasting with the conducting fixed point which is generally unstable to quasiparticle tunneling.

The remainder of this paper is organized as follows. In Sec.~\ref{sec:1}, we introduce the quantum point contact geometry for a general paired FQH edge described by an $so(N)_1 \times u(1)$ conformal field theory, and identify the dominant quasiparticle and electron tunneling processes. In Sec.~\ref{sec:2}, we review the representation theory and primary field content of the $so(N)_1$ sector needed to characterize the neutral tunneling operators. In Sec.~\ref{sec:3}, we analyze weak quasiparticle tunneling using renormalization group arguments, derive the scaling dimension $\Delta_{QP}=(2\nu+N)/8$, and obtain the corresponding power-law corrections to the conductance. In Sec.~\ref{sec:4}, we develop a non-perturbative treatment of the strong quasiparticle tunneling regime via an instanton-gas expansion of the boundary theory, and establish the weak-strong duality that maps this regime to weak electron tunneling, yielding $G\sim T^{2/\nu}$ and $G\sim V^{2/\nu}$ near the insulating fixed point. Finally, in Secs.~\ref{sec:5}, \ref{sec:6}, and \ref{sec:7}, we discuss the resulting transport phenomenology and experimental implications, outline extensions beyond the idealized edge theory, and summarize our conclusions.

\section{Theoretical Model}\label{sec:1}
We consider a paired FQH bar  at filling fraction $\nu$ and composite fermion Chern number $\mathcal{C}_{cf}$. The edge theory is described by one charged chiral boson $\phi$ and $N=|\mathcal{C}_{cf}|$ neutral Majorana fermions which we group as a vector $\Psi = (\psi_1,\cdots,\psi_N)^T$ .  If we ignore possible coupling among the Majoranas or charge-neutral couplings, the neutral part of the edge theory can be described by $N$ free fermions which is equivalent to $so(N)_1$ CFT \cite{yellow_book_cft}. The charged sector can be described by Laughlin-like $u(1)$ CFT. We assume that the FQH bar has infinite boundary condition along the $x$ direction and open boundary condition along the $y$ direction, so that the top and bottom edges host counter-propagating chiral modes. At a QPC located at $x=0$, the two edges are brought into proximity, allowing tunneling processes between them. The Euclidean action for the decoupled top ($t$) and bottom ($b$) edges is given by \cite{PhysRevB.100.035302}:
\begin{align}\label{eq:action_0}
    \mathcal{S}_0 = \sum_{j=t,b} \int d\tau dx \Bigg[ &\frac{1}{4\pi\nu} \partial_x \phi_j \left(i\partial_\tau \phi_j + s_j v_c \partial_x \phi_j\right) \nonumber \\
    &+ \Psi_{j}^T \left(-\partial_\tau + is_j v_n \partial_x\right) \Psi_{j} \Bigg],
\end{align}
where $s_t=+1$ and $s_b=-1$ encode the opposite chiralities, and the velocities of the charge and neutral modes are $v_c$ and $v_n$, respectively with $\mathrm{sgn}(v_n) = \mathrm{sgn}(\mathcal{C}_{cf})$.
\begin{figure}
    \centering
    \includegraphics[width=\linewidth]{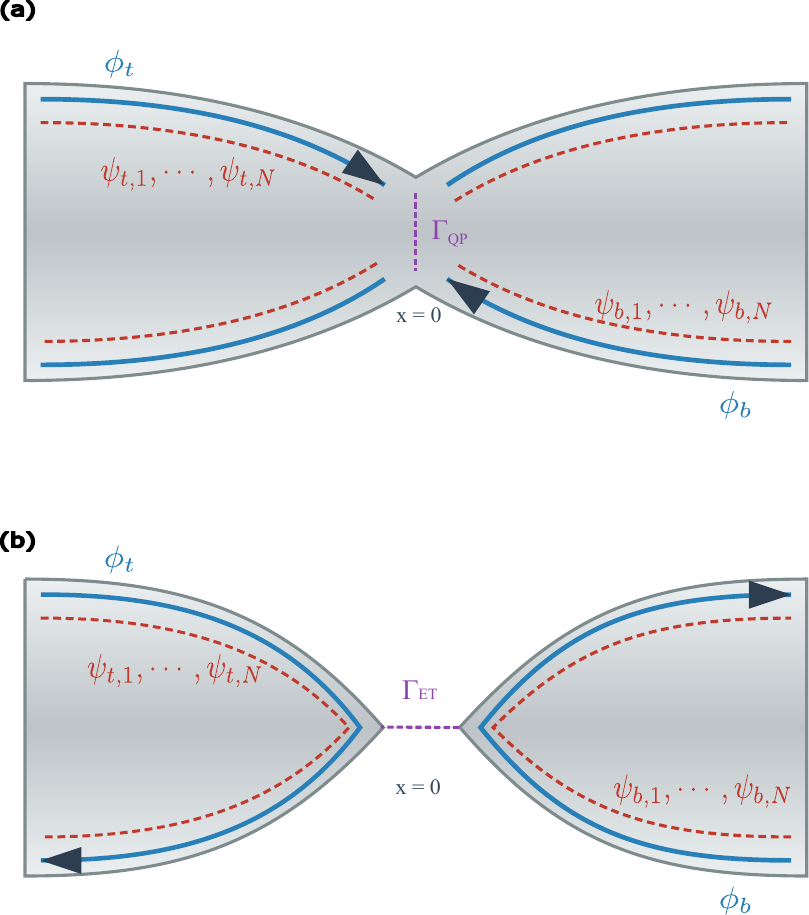}
    \caption{Schematic of a quantum point contact geometry for a paired fractional Quantum Hall state. The top and bottom edges carry a charged boson mode $\phi$ and $N$ neutral Majorana modes $\psi_1,\dots,\psi_N$. The QPC constriction is located at $x=0$. (a) Weak quasiparticle tunneling regime with coupling strength $\Gamma_{QP}$. (b) The electron tunneling regime with coupling strength $\Gamma_{ET}$. }
    \label{fig1}
\end{figure}
At $x=0$, a QPC induces tunneling between the edges. The tunneling Hamiltonian is dominated by the most relevant operator, which corresponds to the transfer of the quasiparticle with the smallest scaling dimension, see Fig.\,\ref{fig1}(a). For paired states, this is the fundamental vortex of the superconductor, carrying charge $e^* = e\nu/2$ and a spin field $\sigma$ from the neutral sector which can be Abelian or non-Abelian depending on $N$. The quasiparticle tunneling at the QPC is given by the following Hamiltonian
\begin{equation}\label{quasiparticle tunneling hamiltonian}
    H_{QPT} = \Gamma_{QP} \Psi^{t\dagger}_{qp}\Psi^b_{qp}+h.c. =\Gamma_{QP} \mathcal{O}_{qp}  \cos\left( \frac{\phi_t -\phi_b}{2} \right)\delta(x)
\end{equation}
where $\Psi_{qp}^j = \sigma^je^{i\phi_j/2}$, $\sigma^j$ is the spin operator of $so(N)_1$ for the $j$-th edge which has $2^{\lfloor N/2\rfloor}$ components, and $\mathcal{O}_{qp}$ is the neutral part of the quasiparticle tunneling Hamiltonian, see Eq.\,\ref{eq: neutral O_qp}.

If the QPC is strong, the FQH bar is effectively cut into left and right FQH bars separated by vacuum. Due to the existence of this vacuum, the only allowed tunneling processes has to involve real particles, or in other words, electron tunneling processes, see Fig.\,\ref{fig1}(b). We can write down the the electron tunneling Hamiltonian of our setup as follows:
\begin{equation}
    H_{ET} = \Gamma_{ET}\Psi_e^{t\dagger}\Psi^b_e+h.c. = i\Gamma_{ET}\Psi^{tT}\Psi^b  \cos\left(\frac{ \phi_t -\phi_b}{\nu} \right)\delta(x)
\end{equation}
where $\Psi^j_e = \Psi_je^{i\phi_j/\nu}$ is the $N$-component electron operator. Our goal is to study charge transport across a QPC in the FQH bar described by the above setup in the weak and strong coupling regimes.

\section{Representation theory of $so(N)$ and Kac-Moody Algebra at level 1} \label{sec:2}
Before we study the charge transport in QPC, it is helpful to understand the underlying algebraic structure of our theory. To this end, we first review the representation theory of $so(N)$. After that, we explain the current algebra and field primaries of $so(N)_1$ CFT.

   \subsection{General properties}
    We know from the theory of Lie Algebras that  $SO(N)$, the rotation group in $N$ dimensions, admits a representation in terms of its universal cover, the spin group $Spin(N)$ \cite{10.1093/acprof:oso/9780199662920.001.0001}. This representation can be efficiently constructed for arbitrary dimension by utilizing Clifford Algebra denoted $Cl(N)$. We define Clifford Algebra $Cl(N)$ by introducing a set of symbols $\Gamma_i$ for $i = 1,\cdots,N$ satisfying the following algebra:
    \begin{equation}
        \Gamma_i\Gamma_j + \Gamma_j\Gamma_i = 2\delta_{ij}
    \end{equation}
    Clifford Algebra is the Algebra generated by $\Gamma_i$ which is a vector space spanned by all possible products of $\Gamma_i$. Note that for $N=3$, the gamma symbols admits a representation in terms of Pauli matrices, while for $N=4$, they are represented by Dirac matrices. In general, the gamma symbols admits a matrix representation in terms of complex-valued matrices of dimension $2^k\times2^k$ where $k=\left \lfloor{\frac{N}{2}}\right \rfloor$. The explicit form of these matrices is of no interest to us and it is sufficient to only consider the algebraic structure of gamma symbols.
    The generators $S_{ij}$ of $so(N)$ can be obtained from Clifford Algebra as follows:
 \begin{equation}
        S_{ij} = -\frac{i}{4} [\Gamma_i, \Gamma_j] = -\frac{i}{2} \Gamma_i \Gamma_j  \quad\quad \text{for } i\neq j
    \end{equation}
    These generators satisfy the commutation relations of the $so(N)$ Lie algebra. In particular, $S_{ij}$ satisfies:
    \begin{equation}\label{eq: lie bracket}
        \comm{S_{ij}}{S_{kl}} = i\left(\delta_{ik}S_{jl}-\delta_{il}S_{jk}-\delta_{jk}S_{il}+\delta_{jl}S_{ik}\right)
    \end{equation}
    Thus, $S_{ij}$ is a representation of the $so(N)$ Lie algebra. Exponentiating these $S_{ij}$, we get two representations, the vector representation of  $so(N)$ and the spinor representation of the universal cover $Spin(N)$. The vector representation corresponds to the adjoint representation under which the $N$ gamma symbols $\Gamma_i$ transform as a vector. Particularly, we have
    \begin{equation}
        U_R\Gamma_iU_R^\dagger = R_{ij}\Gamma_j \quad R\in SO(N),
    \end{equation}
    where $U_R$ is the unitary representation of the rotation matrix $R\in SO(N)$ generated by exponentiating the generators $S_{ij}$. This implies that the gamma symbols serve as basis vectors for a real N-dimensional vector space. In the context of the CFT, these correspond to the $N=|\mathcal{C}_{cf}|$ real Majorana fermion fields $\psi_i(z)$, which have conformal dimension $h_\psi = 1/2$.
    
    On the other hand, the spinor representation is the fundamental representation constructed by acting these generators on a $2^{\lfloor N/2 \rfloor}$-dimensional vector space. The vectors belonging to this vector space are called spinors. For $N=3$, this corresponds to the vector space spanned by spin up and spin down spinors of the electron. These $2^{\lfloor N/2 \rfloor}$ spinors will be promoted to the spin operators in the CFT with conformal dimension $h_\sigma = N/16$. The structure of these spinors depends on the parity of the integers $N$ and $k = \lfloor N/2 \rfloor$.   To classify the different possibilities, we introduce the chirality operator $\Gamma_{\chi} = \prod_{i=1}^{N} \Gamma_i$. The classification is as follows:
    
    \textbf{Odd N:} For odd dimensions ($N = 2k+1$), the chirality operator commutes with all gamma matrices and hence it also commutes with all generators $S_{ij}$. By Schur's lemma, it is proportional to the identity in an irreducible representation. We can confirm that $\Gamma_\chi$ is indeed proportional to the identity. Thus, there is only \textit{one} irreducible spinor representation $\sigma$ of dimension $2^k$.
    
    \textbf{Even N:} For even dimensions ($N = 2k$), the chiral operator $\Gamma_{\chi}$ still commutes with the generators, however, it now anticommutes with the gamma matrices. This allows us to decompose the $2^k$-dimensional spinor space into two irreducible chiral Weyl representations, $\sigma_+$ and $\sigma_-$, each of dimension $2^{k-1}$, distinguished by their eigenvalue under $\Gamma_{\chi}$. Note that we can divide the even $N$ case further depending on the parity of $k=N/2$ as follows:
        \begin{itemize}
            \item Even $k$ (equivalently N = 0 mod 4):
            For this case, we see that $\Gamma_{\chi}\Gamma_{\chi} = +1$. This means that the chirality operator $\Gamma_{\chi}$ has eigenvalues $+1$ or $-1$, implying that the Weyl spinors $\sigma_\pm$ are real or pseudo-real. They are self-conjugate
            \item Odd $k$ (equivalently N = 2 mod 4):
             For this case, we see that $\Gamma_{\chi}\Gamma_{\chi} = -1$. This means that the chirality operator $\Gamma_{\chi}$ has eigenvalues $+i$ or $-i$. This means that the Weyl spinors $\sigma_\pm$ are complex numbers. In particular, $\sigma_\pm$ are complex conjugate of each other.
            
        \end{itemize}

     \subsection{From Algebra to Quantum Operators}
    We now promote these algebraic representations to primary fields in the $so(N)_1$ CFT. The suffix "level 1" indicates that the Kac-Moody algebra is realized by free fermions, restricting the Hilbert space to include only the fundamental integrable representations: the vacuum $I$, the vector $\Psi$, and the spinors $\sigma$ \cite{yellow_book_cft, bernevig2017topological,KITAEV20062, PhysRevB.89.014409, PhysRevB.100.035302}.

    More explicitly, following the discussion in Appendix A of Ref. \cite{Teo2023dihedraltwistliquid}, section 15.5 in Ref. \cite{yellow_book_cft}, and section 3.4 in Ref. in \cite{ralph2009introduction}, we promote the Gamma matrices $\Gamma_i$ to chiral Majorana fermions $\psi_i$, and, we assume that the chiral Majorana fermions have the following correlation function
    \begin{equation}
        \expval{\psi_i(z)\psi_j(z')} = \frac{\delta_{ij}}{z-z'},
    \end{equation}
where $z=e^{2\pi(t+ix)}, z' =e^{2\pi(t'+ix')}$ are complex spacetime coordinates. The Kac-Moody chiral current algebra is generated by the following generators
\begin{equation}
    J_{ij} (z)= -J_{ji} (z)= i:\psi_i\psi_j:(z),
\end{equation}
with $i<j$. These generators span the current algebra of $so(N)_1$. To see that, we compute the operator product expansion of $ J_{ij}J_{kl}$. Using Wick theorem, we find:
\begin{align}
    J_{ij}(z)J_{kl}(0) =&  i(\delta_{jk}J_{il}-\delta_{jl}J_{ik}-\delta_{ik}J_{jl}+\delta_{il}J_{jk})/z \nonumber\\
    &+ (\delta_{il}\delta_{jk}-\delta_{ik}\delta_{jl})/z^2 
\end{align}
Note that the numerator of the first term is nothing but the right hand side of the Lie bracket of $so(N)$. On the other hand, the numerator of the second term is $k=1$ times the killing form $k_{ij,kl}=\Tr{S_{ij}S_{jk}}$. This means that $J_{ij}(z)$ indeed span the current algebra of $so(N)$ at level $k=1$. 

One can check that the Lagrangian of this $so(N)_1$ current algebra constructed using Sugawara Construction matches with the N free Majorana fermion Lagrangian. The primary fields of this theory are obtained by considering the fields that rotate irreducibly under the action of the $so(N)_1$ current operators $J_{ij}(z)$.  We find that these primaries are the vacuum $I$, the vector $\Psi$, and the spinors $\sigma$ for odd $N$ or $\sigma_+$ and $\sigma_-$ for even $N$.

    The conformal weight of the vector field $\Psi$ is $h_\psi = 1/2$. The conformal weight of the spin fields is given by $h_\sigma = N/16$. The fusion rules, which dictate the outcome of bringing two quasiparticles close together, mirror the tensor product decomposition of the group representations, subject to the level truncation.

    \textbf{Odd N:} The fusion algebra contains three primaries: $I, \Psi, \sigma$. The fusion rules are:
    \begin{equation}
        \Psi \times \Psi = I, \quad \Psi \times \sigma = \sigma, \quad \sigma \times \sigma = I + \Psi
    \end{equation}
    We note that despite having multiple Majorana fermions, the fusion algebra is the same as the one of the Ising model with $N=1$.

    \textbf{Even N:} The primaries are $I, \Psi, \sigma_+, \sigma_-$. The fusion of the vector with spinors simply flips the chirality: $\Psi \times \sigma_{\pm} = \sigma_{\mp}$. However, the fusion of two spinors depends on the parity of $k = N/2$:
    \begin{itemize}
        \item \textit{Odd $k$ (e.g., $N=2, 6, 10$):} The Weyl spinors are complex conjugate representations. The fusion rules are:
        \begin{align}
           \Psi \times \Psi = I, \quad \Psi \times \sigma_{\pm} = \sigma_{\mp}, \quad 
           \nonumber\\
           \sigma_\pm \times \sigma_\pm = \Psi, \quad \sigma_+ \times \sigma_- = I
        \end{align}
        Here, two identical spinors fuse to a fermion, while opposite spinors annihilate to the vacuum.
        \item \textit{Even $k$ (e.g., $N=4, 8, 12$):} The Weyl spinors are self-conjugate (real or pseudo-real). The fusion rules are:
        \begin{align}
           \Psi \times \Psi = I, \quad \Psi \times \sigma_{\pm} = \sigma_{\mp}, \quad
           \nonumber\\
           \sigma_\pm \times \sigma_\pm = I, \quad \sigma_+ \times \sigma_- = \Psi
        \end{align}
        In this regime, two identical spinors can annihilate to the vacuum. We note that this set of fusion rules is also known as the toric code fusion rules \cite{KITAEV20062}.
    \end{itemize}
    To end this section, it is worth noting that $so(N)_1$ Kac-Moody current algebra has made several appearances in the wider literature of many-body physics. In particular, the $so(N)_1$ current algebra serves as the basis for the 16-fold way classification of interacting fermions in two dimensions \cite{KITAEV20062, bernevig2017topological} and in paired FQH states \cite{PhysRevB.100.035302}. $so(N)_1$ current algebra was also used in wire construction of twist liquids \cite{Teo2023dihedraltwistliquid} and interacting Dirac nodal superconductors \cite{PhysRevB.98.184514} as well as in constructing generalized spin models with $so(N)_1$ criticality \cite{PhysRevB.89.014409,PhysRevLett.115.237203,PhysRevB.87.041103}.

\section{Renormalization group flow and weak quasiparticle tunneling}\label{sec:3}
\begin{table*}[th!]
\caption{\label{tab:phases} Classification of paired FQH states at filling $\nu=1/2$ within the $so(N)_1 \times u(1)$ framework. The integer $N = |\mathcal{C}_{cf}|$ determines the number of neutral Majorana edge modes. The scaling dimension for quasiparticle tunneling is $\Delta_{QP} = \frac{2\nu+N}{8}$ (see Eq.\,\eqref{quasiparticle tunneling scaling dimension}). For $\nu=1/2$, all listed states are relevant ($\Delta_{QP} < 1$) and flow to the insulating fixed point.}
\begin{ruledtabular}
\begin{tabular}{ccccc}
Chern number ($\mathcal{C}_{cf}$) & State Name & Edge Structure& $\Delta_{QP}$ ($\nu=1/2$) & Relevance \\ \hline
$0$ & Strong Pairing ($K=8$) & Charged Boson & $1/8$& Relevant \\
$+1$ & Moore-Read (Pfaffian) & 1 $\psi$ + Charged Boson & $2/8$& Relevant \\
$-1$ & PH-Pfaffian & 1 $\psi$ + Charged Boson & $2/8$& Relevant \\
$+2$ & Halperin (331) & 2 $\psi$ + Charged Boson & $3/8$& Relevant \\
$-2$ & Halperin (113) & 2 $\psi$ + Charged Boson & $3/8$& Relevant \\
$-3$ & Anti-Pfaffian & 3 $\psi$ + Charged Boson & $4/8$& Relevant \\
$+3$ & $SU(2)_2$ & 3 $\psi$ + Charged Boson & $4/8$& Relevant \\
$-4$ & Anti-(331) & 4 $\psi$ + Charged Boson & $5/8$& Relevant \\
$-5$ & Anti-$SU(2)_2$ & 5 $\psi$ + Charged Boson & $6/8$& Relevant \\
\end{tabular}
\end{ruledtabular}
\end{table*}

Having established the theoretical setup, we now study charge transport in the QPC geometry. First, let us focus on weak quasiparticle tunneling process in Eq.\,\eqref{quasiparticle tunneling hamiltonian}. Using RG analysis, one can easily see that the quasiparticle tunneling strength $\Gamma_{QP}$ follows the following RG equation:
\begin{equation}
    \dfrac{d\Gamma_{QP}}{dl} = (1-\Delta_{QP})\Gamma_{QP}
\end{equation}
where $\Delta_{QP}$ is the scaling dimension of the quasiparticle tunneling Hamiltonian and $e^{-l}=\frac{\Lambda}{\Lambda_0}$ with $\Lambda_0$ and $\Lambda$ being the bare and renormalized energy cutoff \cite{fradkin2013field,PhysRevB.44.5708,PhysRevB.51.2363,fisher1997transport}. The RG equation implies that quasiparticle tunneling is a relevant process for $\Delta_{QP}<1$ and irrelevant for $\Delta_{QP}>1$. The scaling dimension is the sum of the scaling dimensions of the charged and neutral parts of Eq.\,\eqref{quasiparticle tunneling hamiltonian}. We find:
\begin{equation}\label{quasiparticle tunneling scaling dimension}
    \Delta _{QP} = \frac{2\nu+N}{8}
\end{equation}
Since $0<\nu<1$, we see that quasiparticle tunneling is always irrelevant for $N>7$ and always relevant for $N<7$. For $N=7$, quasiparticle tunneling can be relevant, marginal, or irrelevant depending on $\nu$. In Table \,\ref{tab:phases}, we summarize the scaling dimension and relevance of the quasiparticle tunneling for $\nu=\frac{1}{2}$ filling fraction. Similar results can be derived for the recently discovered $\nu = \frac{1}{4},\frac{1}{6},\frac{1}{8},\frac{1}{10}$ fractions \cite{doi:10.1073/pnas.2314212120, PhysRevLett.131.266502,PhysRevLett.134.046502,PhysRevLett.129.156801,kumar2025quarter}. 

We now consider charge transport in the weak quasiparticle tunneling limit following refs. \,\cite{fisher1997transport,PhysRevB.46.15233,Nomura_2001,Ito_2012,PhysRevB.51.2363,PhysRevB.47.3827,PhysRevLett.51.1506,PhysRevB.32.6190,imura1998tunneling}. We apply a dc voltage $V$ at the top left edge and measure the downstream current $I$ across the QPC. To connect the RG result to the experimentally measured scaling behavior of conductance, let $E\equiv \max(k_BT,eV)$ denote the infrared cutoff. Solving the RG equation gives the running tunneling amplitude $\Gamma_{QP}(E)\sim \Gamma_{QP}^{(0)}(E/\Lambda_0)^{\Delta_{QP}-1}$. The leading contribution in perturbation theory to the tunneling current is second order in the tunneling amplitude and has the dimension of energy, so $I(E)\sim |\Gamma_{QP}(E)|^2 E$. Therefore the correction to the ideal open-QPC conductance scales as $\delta G(E)\sim E^{2\Delta_{QP}-2}$. This immediately yields the temperature and voltage exponents quoted below.

When the temperature is larger than the voltage ($k_BT>eV$), $T$ determines the energy cutoff and the tunneling conductance $G$ scales as:
\begin{equation}\label{weak quasiparticle tunneling T>V}
    \nu \frac{e^2}{h}-G \sim  T^{2\Delta_{QP}-2}
\end{equation}
where $\Delta_{QP}$ is the scaling dimension of the quasiparticle tunneling Hamiltonian given in Eq.\,\eqref{quasiparticle tunneling scaling dimension}. For $k_BT<eV$, the voltage $V$ determines the energy cutoff and the tunneling conductance $G$ scales as:
\begin{equation}\label{weak quasiparticle tunneling T<V}
     \nu \frac{e^2}{h} -G \sim  V^{2\Delta_{QP}-2}
\end{equation}
In the low energy regime (low temperature and low voltage regime), the system flows into a stable RG fixed point. For $\Delta_{QP}>1$, the stable fixed point is the conducting fixed point with $\Gamma_{QP}=0$. Hence, we see that weak quasiparticle tunneling transport described in Eqs.\,\eqref{weak quasiparticle tunneling T>V} and \eqref{weak quasiparticle tunneling T<V} corresponds to the low energy regime for $\Delta_{QP}>1$ (conducting fixed point). 

To our knowledge, a conducting stable fixed point has never been experimentally observed. Generally speaking, we believe that high $N$ FQH states are less likely to develop and are likely unstable. Thus, we see that for experimentally feasible scenarios, quasiparticle tunneling is relevant and will always flow to the strong quasiparticle tunneling regime at low voltage (low temperature) and a non-perturbative treatment is needed. In the next section, we provide a non-perturbative analysis based on instanton gas approximation \cite{fisher1997transport,PhysRevB.46.15233,Nomura_2001,Ito_2012,PhysRevB.51.2363,PhysRevB.47.3827,PhysRevLett.51.1506,PhysRevB.32.6190}.
\section{Strong quasiparticle tunneling}\label{sec:4}
Having established the weak quasiparticle tunneling regime, we now consider the strong quasiparticle tunneling regime which corresponds to the $\Gamma_{QP}=\infty$ fixed point. To that end, we generalize the instanton gas approximation originally developed for the Pfaffian state \cite{Nomura_2001}, the 331 state \cite{Nomura_2001}, and the anti-Pfaffian state \cite{Ito_2012} to the general case described by (1+1)d chiral $so(N)_1\otimes u(1)$ CFT.  
To facilitate the instanton analysis, we bosonize the neutral sector of the   $so(N)_1\otimes u(1)$ CFT. To do this, we need to combine two Majorana fermions with the same chirality to form a single chiral Dirac fermion. Then, using abelian bosonization, we express the Dirac fermion as an exponential of a chiral boson. One way to do this is by grouping pairs of Majoranas on the same edge. This is possible only for even $N$. if $N$ is odd, we will always be left with a single unpaired Majorana which, at first glance, implies that the bosonization approach is hopeless for odd $N$. In this work, we use an alternative approach which puts even and odd $N$ at equal footing.

Following ref \cite{PhysRevLett.97.036801}, one can flip the chirality of the bottom edge. This process allows us to have $2N$ co-propagating Majorana fermions that we can always pair together. This chirality flipping process is defined by introducing the following fields:
\begin{subequations}
    \begin{equation}
        \phi_1 (\tau,x) = \phi_t(\tau,x), \quad \psi_{2j-1}(\tau,x) = \psi_j^t(\tau,x) 
    \end{equation}
    \begin{equation}
        \phi_2(\tau,x) = \phi_b(\tau,-x) , \quad \psi_{2j}(\tau,x) = \psi_j^b(\tau,-x) 
    \end{equation}
\end{subequations}
Now, we can pair the Majoranas $\psi_{2j-1}$ and $\psi_{2j}$ (where $j=1,\dots,N$) to form Dirac fermions $\psi_D^{j}$:
\begin{equation}
    \psi_D^{j} = \psi_{2j-1} + i \psi_{2j} \sim  e^{i\chi_{j}} 
\end{equation}
Here, $\chi_{j}$ are chiral bosonic fields satisfying the standard commutation relations. Their free Lagrangian is equivalent to that of $\nu=N$ edge. 

Bosonizing the spinor operator $\sigma$ is rather tricky especially for odd $N$. In particular, as discussed in Sec. \ref{sec:2}, we have $2^{\lfloor N/2 \rfloor}$ possible spinor operators with non-trivial operator product expansion. Rather than bosonizing the spinor operators, we bosonize the neutral part of the quasiparticle tunneling Hamiltonian instead. Under mild assumptions of Hermiticity and equal likelihood of tunneling between the $2^{\lfloor N/2 \rfloor}$ possible spinors, the neutral part of the quasiparticle tunneling Hamiltonian takes the following general form \cite{DIFRANCESCO1987527,PhysRevB.75.045317}:

\begin{equation}\label{eq: neutral O_qp}
    \mathcal{O}_{QP} \sim \prod_{j=1}^N\cos\left(\chi_j/2\right)
\end{equation}
In terms of this new set of fields, and up to possible unimportant total derivatives, we rewrite the Lagrangian in bosonic form as follows:
\begin{align}
    \mathcal{L}_{bosonized} = &\frac{1}{4\pi\nu}\sum_{j=1,2}   \partial_x \phi_j\left(i\partial_\tau \phi_j  + v_c \partial_x \phi_j\right) 
    \nonumber
    \\
    &+ \frac{1}{4\pi}\sum_{j=1}^N   \partial_x \chi_j\left(i\partial_\tau \chi_j  + v_n \partial_x \chi_j\right)
    \nonumber
    \\
    &- \Gamma_{QP}\cos\left(\frac{\phi_1-\phi_2}{2}\right)\prod_{j=1}^N\cos\left(\chi_j/2\right)\delta(x)
    \label{bosonized lagrangian}
\end{align}
We now construct an effective field theory for the boundary fields at the quantum point contact ($x=0$). We define the boundary fields $\varphi(\tau) \equiv \phi_1(\tau, 0) -\phi_2(\tau, 0)$ and $\vartheta_j(\tau) \equiv \chi_j(\tau, 0)$. To this end, we consider the partition function of the bosonized Lagrangian \eqref{bosonized lagrangian}
:
\begin{align}
    Z &= \int\mathcal{D}\phi_1\mathcal{D}\phi_2\prod_j^N\mathcal{D}\chi_j \exp(-S_{bosonized})
    \nonumber
    \\
    &= \int\mathcal{D}\phi_1\mathcal{D}\phi_2\mathcal{D}\varphi\prod_j^N\mathcal{D}\chi_j\mathcal{D}\vartheta_j \delta(\varphi-(\phi_1-\phi_2)|_{x=0})
    \nonumber
    \\
    &\times \delta(\vartheta_j-\chi_j|_{x=0}) \exp(-\tilde{S}_{bosonized})
    \nonumber
    \\
&=\int\mathcal{D}\phi_1\mathcal{D}\phi_2\mathcal{D}\varphi\mathcal{D}\lambda_0\prod_j^N\mathcal{D}\chi_j\mathcal{D}\vartheta_j\mathcal{D}\lambda_j  \exp(-\tilde{S}_{bosonized})
\nonumber
\\
&\times \exp(-\int\dd \tau \left[\lambda_0\left(\varphi-\left(\phi_1-\phi_2\right)|_{x=0}\right)+\lambda_j\left(\vartheta_j-\chi_j|_{x=0}\right)\right])
\end{align}
where $S_{bosonized}=\int\dd \tau\dd x \mathcal{L}_{bosonized}$ and $\tilde{S}_{bosonized}=\int\dd \tau\dd x \mathcal{\tilde{L}}_{bosonized}$
where $\mathcal{\tilde{L}}_{bosonized}$ is the bosonized Lagrangian with the substitution $\varphi(\tau) \equiv \phi_1(\tau, 0) -\phi_2(\tau, 0)$ and $\vartheta_j(\tau) \equiv \chi_j(\tau, 0)$ in the cosine term.  Integrating out the fields $\phi_{1,2},\chi_j,\lambda_j$ yield the following partition function
\begin{equation}
    Z = \int\mathcal{D}\varphi\prod_j\mathcal{D}\vartheta_j\exp(-S_{eff})
\end{equation}
where $S_{eff}$ is the effective boundary action. The free part of the action is
\begin{equation}\label{S_eff_0}
    S_{eff}^{(0)} = \frac{1}{4\pi \nu} \sum_{\omega} |\omega| |\varphi(\omega)|^2 + \frac{1}{4\pi} \sum_{j=1}^{N} \sum_{\omega} |\omega| |\vartheta_j(\omega)|^2
\end{equation}

We now consider the tunneling term. The total effective action in the strong tunneling regime is:
\begin{equation}\label{S_strong}
    S_{eff} = S_{eff}^{(0)} - \Gamma_{QP} \int d\tau \cos\left(\frac{\varphi(\tau)}{2}\right) \prod_{j=1}^{N} \cos\left(\frac{\vartheta_j(\tau)}{2}\right)
\end{equation}

We analyze the strong coupling limit $\Gamma_{QP} \to \infty$. In this regime, the fields $\varphi$ and $\vartheta_j$ are pinned to the minima of the potential. We treat the transport perturbatively by considering tunneling events (instantons) between these minima.

Using trigonometric identity, the potential is
\begin{align}
    V(\varphi, \vartheta) &= -\cos(\varphi/2)\prod_j \cos(\vartheta_j/2)
    \nonumber
    \\
    &\propto -\sum_{q_j=\pm} \cos(\frac{\varphi+q_1\vartheta_1+\cdots+q_N\vartheta_N}{2})
\end{align}
This potential is minimized when the total arguments inside the cosine terms is a multiple of $2\pi$. Alternatively, it is minimized when an even number of boundary bosonic fields equal to $2\pi n$ for some integer $n$.

Let us analyze the hopping of instantons between different minima of the potential. The fundamental charged excitation corresponds to a $2\pi$ phase slip in the relative charge phase $\varphi$:
\begin{equation}
    \Delta \varphi = 2\pi
\end{equation}
Substituting this jump into the potential term:
\begin{equation}
    \cos\left(\frac{\varphi + 2\pi}{2}\right) = \cos\left(\frac{\varphi}{2} + \pi\right) = -\cos\left(\frac{\varphi}{2}\right)
\end{equation}
The charge term flips sign. To maintain the vacuum energy (i.e., to keep the potential $V(\varphi, \vartheta)$ negative), the neutral sector product $\prod_{j=1}^M \cos(\vartheta_j/2)$ must \textit{also} flip sign.

To flip the sign of the product $\prod_{j=1}^M \cos(\vartheta_j/2)$, we must shift an odd number of the fields $\vartheta_j$ by $2\pi$. Since the action cost of an instanton scales with the square of the jump (Eqs.\,\eqref{S_eff_0},\eqref{S_strong}), the dominant contribution comes from the minimal jump. In other words, the fundamental instanton involves a shift of $\Delta \vartheta_k = 2\pi$ for exactly one neutral flavor $k \in \{1, \dots, N\}$, and zero for all others ($j \neq k$). Thus, there are $N$ distinct types of fundamental instantons, labeled by the neutral flavor index $k$.

We approximate the field configuration as a gas of $N_{inst}$ instantons at times $\tau_i$ with charges $q_i,\tilde{q}_i = \pm 1$ and flavor indices $p_i$.
\begin{equation}
     \varphi(\tau) = \sum_i q_i 2\pi \Theta(\tau - \tau_i), \quad \vartheta_{p_i}(\tau) = \sum_i \tilde{q}_i 2\pi \Theta(\tau - \tau_i) \delta_{p, p_i}
\end{equation}
where $\Theta$ is the Heaviside step function. The Fourier transform of a step function is $2\pi / (i\omega)$. The above expression minimizes the potential $V(\varphi, \vartheta)$. In the strong quasiparticle tunneling regime $\Gamma_{QP}$, we can treat the kinetic action $S_{eff}^{(0)}$ as a perturbation. Substituting the above field configuration into the kinetic action $ S_{eff}^{(0)}$, we get the instanton gas action. For the charged sector, we get
\begin{align}
    S_{inst}^{\varphi} &= \frac{1}{4\pi\nu} \int \frac{d\omega}{2\pi} |\omega| \left| \sum_j \frac{2\pi q_j}{\omega} e^{i\omega \tau_j} \right|^2 \nonumber \\
    &= \frac{\pi}{\nu} \sum_{i,j} q_i q_j \int \frac{d\omega}{2\pi} \frac{1}{|\omega|} e^{i\omega(\tau_i - \tau_j)}
\end{align}
Using the regularization $\int \frac{d\omega}{|\omega|} e^{i\omega \tau} \sim -2 \ln(\tau/\tau_c)$, we obtain the logarithmic interaction:
\begin{equation}
    S_{inst}^{\varphi} = -\frac{1}{\nu} \sum_{i \neq j} q_i q_j \ln\left(\frac{|\tau_i - \tau_j|}{\tau_c}\right) + \cdots
\end{equation}
For the neutral field $\vartheta_p$, the contribution is non-zero only if instantons $i$ and $j$ share the same flavor $p$.
\begin{equation}
    S_{inst}^{\vartheta} = -1 \sum_{i \neq j} \delta_{p_i, p_j} \tilde q_i \tilde q_j \ln\left(\frac{|\tau_i - \tau_j|}{\tau_c}\right) + \cdots
\end{equation}

The partition function for the instanton gas is:
\begin{align}
    Z =& \sum_{n} \sum_{\{p_i\}} \frac{y^{2n}}{(n!)^2} 
    \nonumber
    \\
    &\int \prod d\tau_i \exp\left[ \sum_{i \neq j} \left(\frac{q_i q_j}{\nu} +\tilde q_i \tilde q_j \delta_{p_i, p_j}\right)  \ln|\tau_{i}-\tau_j| \right]
\end{align}
This represents a multi-component Coulomb gas.
Note that this partition function is equivalent to that of a sine-Gordon theory. In particular, each term in the above summation is equivalent to a perturbative expansion of the potential term in sine-Gordon theory with coupling constant $y$.

We map this Coulomb gas back to a field theory using the standard duality between the Coulomb gas and the Sine-Gordon model. We introduce dual fields $\tilde{\varphi}$ and $\tilde{\vartheta}_p$. The logarithmic interaction with coefficient $K_{eff}\equiv \frac{1}{\nu}+1$ maps to a cosine operator with scaling dimension $K_{eff}$ \cite{PhysRevE.56.619,PhysRevLett.51.1506,narayan19992d}.

Specifically, an instanton with charge jumps $(\Delta \varphi, \Delta \vartheta_p) = (2\pi, 2\pi)$ corresponds to the vertex operator:
\begin{equation}
    \mathcal{O}_{dual}^{(p)} \sim \exp\left( i \tilde{\varphi}/\nu + i \tilde{\vartheta}_p \right)
\end{equation}
Here, the dual fields are normalized such that $e^{i\tilde{\varphi}/\nu}$ creates a $2\pi$ kink in $\varphi$.
 Physically, a $2\pi$ kink in the quasiparticle phase $\varphi$ corresponds to the tunneling of a full electron charge $e$.
A $2\pi$ kink in the neutral boson $\tilde\vartheta_p$ corresponds to the Dirac fermion mass operator for the $p$-th flavor, $\psi_{Dp}^\dagger \psi_{Dp}=i\psi_{2p-1}\psi_{2p}=i\psi^t_p\psi^b_p$. 

The effective dual action is therefore:
\begin{equation}\label{dual effective action}
    S_{dual} = S_{0,dual}-y \sum_{k=1}^{N}\int d\tau   \cos\left( \tilde{\varphi}/\nu \right)\cos\left( \tilde{\vartheta}_k \right) 
\end{equation}
where $y$ is the fugacity of the instanton (tunneling amplitude) and $S_{0, dual}$ is given by:
\begin{equation}
    S_{0, dual} = \frac{1}{4\pi \nu} \sum_{\omega} |\omega| |\tilde\varphi(\omega)|^2 + \frac{1}{4\pi} \sum_{j=1}^{N} \sum_{\omega} |\omega| |\tilde\vartheta_j(\omega)|^2
\end{equation}
Comparing Eq.\,\eqref{dual effective action} to Eq.\,\eqref{S_strong}, we see that the dual theory including bulk degrees of freedom is simply the original problem with the tunneling term replaced with weak electron tunneling process. Finally, we obtain:
\begin{align}\label{dual lagrangian}
    \mathcal{L}_{dual} &= \sum_{j=t,b} \Big[ \frac{1}{4\pi\nu} \partial_x \phi_j\left(i\partial_\tau \phi_j  + s_jv_c \partial_x \phi_j\right) 
    \nonumber
    \\
    &+ \Psi_{j}^T \left(-\partial_\tau + is_j v_n \partial_x\right) \Psi_{j} \Big] - H_{ET}
\end{align}
with
\begin{equation}
    H_{ET} = iy\Psi_t^T\Psi_b\cos\left(\frac{\phi_t-\phi_b}{\nu}\right)\delta(x)
\end{equation}
Here, $s_t=1$, and  $s_b=-1$.

Having established the dual theory, we now discuss the charge transport in the strong quasiparticle tunneling regime. Electron tunneling of the dual theory has the following RG equation:
\begin{equation}
    \frac{dy}{dl} = (1-\Delta_{ET})y
\end{equation}
where $\Delta_{ET}$ is the the scaling dimension of the electron tunneling operator. Since the neutral part of the electron tunneling is a summation of Majorana fermion bilinear terms for all $N$, we see that the scaling dimension of the electron tunneling Hamiltonian is the same for all competing states and is given by:
\begin{equation}
    \Delta_{ET} = \frac{1}{\nu} +1
\end{equation}
Since $\nu<1$, electron tunneling is an irrelevant operator and can be treated perturbatively. The derivation of the scaling law of conductance in the dual theory parallels the weak quasiparticle tunneling case. If $E\equiv \max(k_BT,eV)$, the instanton fugacity obeys $y(E)\sim y_0(E/\Lambda_0)^{\Delta_{ET}-1}$. The transmitted current through the pinched-off QPC is quadratic in this running coupling, $I(E)\sim |y(E)|^2E$, so the conductance scales as $G(E)\sim E^{2\Delta_{ET}-2}$. Because the neutral Majorana bilinear contributes scaling dimension $1$ while the charge vertex contributes $1/\nu$, the exponent is independent of $N$ and depends only on the filling fraction. Thus we obtain for $k_BT>eV$, the tunneling conductance $G$ scales as:
\begin{equation}\label{weak electron tunneling T>V}
    G \sim  T^{2\Delta_{ET}-2} = T^{2/\nu}
\end{equation}
For $k_BT<eV$, the tunneling conductance $G$ scales as:
\begin{equation}\label{weak electron tunneling T<V}
     G \sim  V^{2\Delta_{ET}-2} = V^{2/\nu}
\end{equation}
The above two equations describe the conductance near zero voltage/zero temperature in the insulating phase (small number of Majorana $N$). Alternatively, it describes the high voltage/high temperature regime in the conducting phase (large number of Majorana $N$).
\begin{table}[th!]
\centering
\begin{tabular}{llc}
\toprule
Energy regime & Condition & Conductance \\
\midrule
\multirow{2}{*}{Low energy}
 & $V > T$ 
 & $G(V) \propto V^{2/\nu}$ \\
 & $T > V$ 
 & $G(T) \propto T^{2/\nu}$ \\
\midrule
\multirow{2}{*}{High energy}
 & $V > T$ 
 & $\nu \dfrac{e^2}{h} - G(V) \propto V^{2\Delta_{\mathrm{QP}}-2}$ \\
 & $T > V$  
 & $\nu \dfrac{e^2}{h} - G(T) \propto T^{2\Delta_{\mathrm{QP}}-2}$ \\
\bottomrule
\end{tabular}
\caption{Scaling behavior of the differential conductance $G$ in different energy regimes as a function of voltage $V$ and temperature $T$ for paired FQH states with small number of Majorana fermions $(\Delta_{\mathrm{QP}}<1)$.}
\label{tab:scaling_conductance}
\end{table}

\section{Discussion}\label{sec:5}
The results derived in Sections \ref{sec:3} and \ref{sec:4} allow us to construct a comprehensive phase diagram for charge transport in paired FQH states. The transport phenomenology is strictly dictated by the renormalization group (RG) flow of the quasiparticle tunneling amplitude $\Gamma_{QP}$. The direction of this flow is controlled by the scaling dimension
\begin{equation}
\Delta_{QP}=\frac{2\nu+N}{8},
\end{equation}
where $N=|\mathcal{C}_{cf}|$ is the number of neutral Majorana modes. Depending on whether quasiparticle tunneling is relevant or irrelevant, the system flows to distinct infrared (IR) fixed points, leading to qualitatively different low-energy transport behavior.

When $\Delta_{QP}<1$, quasiparticle tunneling is relevant and grows under RG. In this case, the system flows in the IR ($T,V\to0$) to the strong quasiparticle tunneling (pinched-off QPC) fixed point, and the conductance vanishes. The low-energy transport is then controlled by the dual weak electron tunneling description, yielding the universal scaling
\begin{equation}
G(T)\propto T^{2/\nu}, \qquad G(V)\propto V^{2/\nu},
\end{equation}
which depends only on the filling fraction $\nu$ and is independent of $N$. Thus, for experimentally relevant paired states with small $N$ (Pfaffian, 331, 113, anti-Pfaffian, PH-Pfaffian), the low-temperature and low-bias response is universally insulating.

At higher temperature or voltage, the system crosses over toward the unstable open-QPC fixed point. In this ultraviolet (UV) regime, the conductance approaches the quantized plateau $\nu e^2/h$, and the leading deviation is governed by weak quasiparticle tunneling,
\begin{equation}
\nu\frac{e^2}{h}-G(T)\propto T^{2\Delta_{QP}-2}, \qquad
\nu\frac{e^2}{h}-G(V)\propto V^{2\Delta_{QP}-2}.
\end{equation}
Crucially, this exponent retains explicit dependence on $N$, providing a direct electrical fingerprint of the neutral Majorana sector. Therefore, for small $N$, the universal behavior appears in the IR, while the $N$-dependent scaling is visible only in the high-energy approach to the quantized plateau. We summarize these results in Table.\,\ref{tab:scaling_conductance}

In actual experiments for arbitrary $N$, both the weak and strong quasiparticle tunneling regimes will be observed in experiment as one increases the temperature/applied voltage. One interpolates between these two regimes and the crossover energy scale separating these two regimes is given by \cite{imura1998tunneling}
\begin{equation}
    T_{crossover}, V_{crossover} \sim \Gamma_{bare}^{1/(1-\Delta_{QP})}
\end{equation}
where $\Gamma_{bare}$ is the bare coupling strength of the quasiparticle tunneling. The bare coupling strength is controlled by the gate voltage at the QPC. To minimize the crossover energy scale and make it easier to observe the scaling behavior associated with weak quasiparticle tunneling, a small gate voltage is necessary. This allow us to minimize the parameter regime of the universal infrared behavior while maximizing the parameter regime of the $N$-dependent ultraviolet scaling, allowing us to directly infer the number of Majorana modes.

\section{Outlook}\label{sec:6}
Our results can be extended in several directions. Throughout this work we assumed the minimal unreconstructed edge, consisting of one charged mode and $N$ neutral Majorana modes, and neglected additional charge-neutral or neutral-neutral couplings. In realistic devices, however, a smooth confining potential can reconstruct the edge and generate extra counter-propagating charged or neutral modes. Once such extra modes are present, Coulomb interactions and random impurity scattering can drive equilibration and may lead to disorder-dominated Kane-Fisher-Polchinski--like fixed points with non-universal conductance exponents \cite{kane1994randomness, PhysRevLett.132.136502,p4y5-trph,parkFingerprintsAntiPfaffianTopological2024,asasiEquilibrationEdgeStates2021,spanslattContactsEquilibrationInteractions2021,dasEffectInteredgeCoulomb2009}. For purely chiral minimal edges ($\mathcal{C}_{cf}\geq0$), disorder is less effective because all modes co-propagate, although soft confinement can still create reconstructed edges. By contrast, non-chiral or particle-hole-conjugate paired states  ($\mathcal{C}_{cf}<-1$) are more susceptible to equilibration and can in principle exhibit intermediate plateaus, including  $G=e^2/h$ scenario predicted for anti-Pfaffian \cite{Ito_2012,spanslattContactsEquilibrationInteractions2021}. The present framework can be extended to such situations by enlarging the edge theory to include the reconstructed modes and then repeating the same boundary RG and instanton analysis for the resulting coupled boson-Majorana or more general $K$-matrix theory.

Another important aspect is we assumed uniform coupling strength for all possible quasiparticle tunneling. In reality, we have $2^{\lfloor N/2 \rfloor}$ possible quasiparticles and we can assign different coupling strength for each quasiparticle tunneling process. We expect that intermediate phases can appear especially for the case with $\mathcal{C}_{cf}<-1$ \cite{Ito_2012,imura1999theory}.

Furthermore, our results provide a universal starting point for a general transport framework across quantum point contacts and interfaces separating arbitrary topological edges, including next generation paired states \cite{yutushuiTheoryNextGenerationEvenDenominator2026} or paired states coupled to generic Abelian $K$-matrix states. In particular, QPC geometry separating paired FQH state and Integer Quantum Hall state where Andreev-like reflection process is possible\cite{ohashi2022andreev}.

\section{Conclusion}\label{sec:7}
We have developed a unified transport theory for paired FQH states across a quantum point contact. By employing the instanton approximation, we demonstrated a duality between strong quasiparticle tunneling and weak electron tunneling.

We developed a unified transport theory for quantum point contacts in paired even-denominator fractional quantum Hall states whose low-energy edge physics is described by an $so(N)_1\times u(1)$ conformal field theory. The integer $N=|\mathcal{C}_{cf}|$ encodes the number of neutral Majorana modes and distinguishes competing paired topological orders.

We identified the fundamental quasiparticle tunneling process carrying charge $e^\ast=e\nu/2$ and derived its scaling dimension $\Delta_{QP}=(2\nu+N)/8$. Depending on whether this operator is relevant or irrelevant, the system flows in the infrared to either a universally insulating pinched-off fixed point or a conducting open-QPC fixed point. For small $N$, quasiparticle tunneling is relevent. In the stable infrared regime, the nonlinear conductance exhibits universal scaling $G\sim T^{2/\nu}$ or $G\sim V^{2/\nu}$, independent of $N$. In contrast, the unstable fixed point displays $N$-dependent power-law corrections $G\sim T^{2\Delta_{QP}-2}$ or $G\sim V^{2\Delta_{QP}-2}$, providing a direct electrical fingerprint of the neutral Majorana sector.

Using a non-perturbative instanton expansion, we established an exact weak--strong duality mapping strong quasiparticle tunneling to weak electron tunneling, placing the transport problem for all paired states within a single controlled framework.

Together, these results organize the transport phenomenology of paired fractional quantum Hall states into two universal asymptotic regimes connected by a crossover whose scaling encodes the underlying topological order. They provide a practical and purely electrical route to identifying Majorana content in even-denominator quantum Hall states, complementing thermal transport and interferometric probes.
\begin{acknowledgments}
    E.A. K.N, and Y.T. acknowledge support from JSPS with Grants-in-Aid for Scientific research (KAKENHI Grant No. 25H00613 ). K.N. is supported in part by No. JP25H01250.
    R.O. is supported by JSPS KAKENHI Grant No. JP25K23361. 
    H.I. is supported in part by JSPS KAKENHI Grant No. JP24H00197. 
\end{acknowledgments}

\bibliography{biblio}
\end{document}